\documentclass[prd,twocolumn,superscriptaddress,altaffilletter,showpacs,nofootinbib]{revtex4}
\usepackage{amssymb}
\usepackage[dvips]{graphicx}
\usepackage{amsmath}

\newcommand{\be}{\begin{equation}}
\newcommand{\ee}{\end{equation}}
\newcommand{\bea}{\begin{eqnarray}}
\newcommand{\eea}{\end{eqnarray}}
\newcommand{\bib}{\bibitem}
\newcommand{\der}{\partial}
\newcommand{\vphi}{\varphi}
\newcommand{\n}{\nabla}

\newcommand{\lab}{\label}

\begin{document}

\title{The conformal transformation's controversy: what are we missing?}

\author{Israel Quiros}\email{iquiros6403@gmail.ugto.mx}\affiliation{Departamento de Matem\'aticas, Centro Universitario de Ciencias Ex\'actas e Ingenier\'{\i}as (CUCEI), Corregidora 500 S.R., Universidad de Guadalajara, 44420 Guadalajara, Jalisco, M\'exico.}

\author{Ricardo Garc\'{\i}a-Salcedo}\email{rigarcias@ipn.mx}\affiliation{Centro de Investigacion en Ciencia Aplicada y Tecnologia Avanzada (CICATA), Legaria del IPN, M\'exico D.F., M\'exico.}

\author{Jose Edgar Madriz Aguilar}\email{madrizaguilar@yahoo.com.mx}\affiliation{Departamento de Matem\'aticas, Centro Universitario de Ciencias Ex\'actas e Ingenier\'{\i}as (CUCEI), Corregidora 500 S.R., Universidad de Guadalajara, 44420 Guadalajara, Jalisco, M\'exico.}

\author{Tonatiuh Matos}\email{tmatos@fis.cinvestav.mx}\affiliation{Departamento de F{\'\i}sica, Centro de Investigaci\'on y de Estudios Avanzados del IPN, A.P. 14-740, 07000 M\'exico D.F., M\'exico.}

\date{\today}

\begin{abstract}
An alternative interpretation of the conformal transformations of the metric is discussed according to which the latter can be viewed as a mapping among Riemannian and Weyl-integrable spaces. A novel aspect of the conformal transformation's issue is then revealed: these transformations relate complementary geometrical pictures of a same physical reality, so that, the question about which is the physical conformal frame, does not arise. In addition, arguments are given which point out that, unless a clear statement of what is understood by ''equivalence of frames'' is made, the issue is a semantic one. For definiteness, an intuitively ''natural'' statement of conformal equivalence is given, which is associated with conformal invariance of the field equations. Under this particular reading, equivalence can take place only if the metric is defined up to a conformal equivalence class. A concrete example of a conformal-invariant theory of gravity is then explored. Since Brans-Dicke theory is not conformally invariant, then the Jordan's and Einstein's frames of the theory are not equivalent. Otherwise, in view of the alternative approach proposed here, these frames represent complementary geometrical descriptions of a same phenomenon. The different points of view existing in the literature are critically scrutinized on the light of the new arguments.
\end{abstract}

\pacs{02.40.-k, 02.40.Ky, 02.40.Hw, 04.20.-q, 04.20.Cv, 04.50.Kd, 04.50.+h, 11.25.Wx}
\maketitle

\section{Introduction}\lab{intro}

Brans-Dicke (BD) theory of gravity \cite{bd} represents the most simple generalization of general relativity (GR). The theory is parametrized by a constant parameter $\omega$ -- the BD coupling. Up to a boundary term in the action the vacuum BD field equations can be derived from:

\be S_{BD}=\frac{1}{16\pi}\int d^4x\sqrt{-g}\;e^\vphi\left[R-\omega (\n\vphi)^2\right],\lab{bd-action}\ee where $R$ is the curvature scalar, $(\n\vphi)^2\equiv g^{\mu\nu}\n_\nu\vphi\n_\mu\vphi$, and we have introduced the BD scalar field $\vphi$ which is related to the original one \cite{bd} through $\phi=e^\vphi$. The derived field equations -- which model the BD laws of gravity -- are

\bea &&G_{\mu\nu}=\omega\left[\der_\mu\vphi\der_\nu\vphi-\frac{1}{2}g_{\mu\nu}(\n\vphi)^2\right]+\der_\mu\vphi\der_\nu\vphi\nonumber\\
&&\;\;\;\;\;\;\;\;\;\;\;\;\;\;\;\;\;\;\;\;-g_{\mu\nu}(\n\vphi)^2+\n_\mu\n_\nu\vphi-g_{\mu\nu}\Box\vphi,\lab{v-bd-feqs}\eea where $\Box\equiv g^{\mu\nu}\n_\mu\n_\nu$, and the Klein-Gordon (KG) equation for the scalar field:

\be (2\omega+3)[\Box\vphi+(\n\vphi)^2]=0.\lab{v-bd-kg-eq}\ee 

The theory can be formulated in different sets of field variables \cite{d-units}, among which we may cite BD gravity in Jordan's frame (JF) variables -- the standard formulation of the theory \cite{bd} given by equations (\ref{bd-action}), (\ref{v-bd-feqs}), (\ref{v-bd-kg-eq}) -- and BD theory in the so called Einstein's frame (EF) \cite{d-units}, which can be obtained from the Jordan frame formulation by a conformal transformation of the metric $\bar g_{\mu\nu}=e^\vphi g_{\mu\nu}$. For vacuum, up to a boundary term in the action, the latter formulation of the theory can be derived from,

\be \bar S_{BD}=\frac{1}{16\pi}\int d^4 x\sqrt{-\bar g}\left[\bar R-\left(\omega+\frac{3}{2}\right)(\bar\n\vphi)^2\right],\lab{ef-bd-action}\ee where the over-bar means the quantities are given in terms of the conformal metric $\bar g_{\mu\nu}$. The derived field equations, plus the KG equation are, 

\bea &&\bar G_{\mu\nu}=\left(\omega+\frac{3}{2}\right)\left[\bar\der_\mu\vphi\bar\der_\nu\vphi-\frac{1}{2}\bar g_{\mu\nu}(\bar\n\vphi)^2\right],\lab{efbd-feqs}\eea and 

\be \bar\Box\vphi=0,\lab{efbd-kg-eq}\ee respectively. Equivalence of JF and EF formulations of BD theory of gravity under conformal transformations of the metric, also known as transformations of units \cite{d-units},

\be \bar g_{\mu\nu}=\Omega^2 g_{\mu\nu},\lab{c-t}\ee where $\Omega^2=e^\vphi$, has been discussed in the literature since long ago \cite{d-units,fierz}, and more recently has been put to discussion again in connection with scalar-tensor theories of gravity \cite{magnano,capozziello-cqg-1997,faraoni-rev-1999,faraoni-ijtp-1999,vollick,flanagan,faraoni-prd-2007,catena}. In spite of the amount of work published on this subject to date (see the review \cite{faraoni-rev-1999}), the controversy is still open \cite{capozziello-plb-2010,capozziello-mpla-2010,deruelle-sasaki,corda}. The discussion in the following paragraphs will support this statement.\footnote{Needless to say that the conformal transformation's issue is critical for the interpretation of the predictions of given scalar-tensor theories of gravity since these are deeply affected by the choice of the conformal frame \cite{magnano,faraoni-rev-1999,vollick,flanagan,corda}. It is of central importance also for the understanding of the physics behind the graviton-dilaton string effective theory \cite{wands} since, independent of the dimensionality of the spacetime and the number of compactified dimensions, the string frame (SF) dilaton-gravity action is nothing but JFBD action with, $\omega=-1$ (see, however, Ref.\cite{ruso}). The string effective theory may be formulated in a number of conformal frames as well, including the SF and the EF among others.}

Even if there can be different points of view on the subject, it seems intuitively ''natural'' to associate conformal equivalence with invariance under the conformal transformations (\ref{c-t}). If one follows this intuitive understanding of conformal equivalence, since BD theory is not conformal-invariant, then there can not be dynamical equivalence among the Jordan's and Einstein's frames. This is easily demonstrated in the simplest case of vacuum theory. In fact, under a conformal transformation (\ref{c-t}), with $\Omega^2=e^\vphi$, the vacuum JFBD action (\ref{bd-action}) is mapped into the EFBD action (\ref{ef-bd-action}), while the JF vacuum field equations (\ref{v-bd-feqs}), (\ref{v-bd-kg-eq}), are transformed into the vacuum field equations in the EF: Eq.s (\ref{efbd-feqs}), and (\ref{efbd-kg-eq}). I. e., the laws of gravity -- expressed through the field equations -- are not invariant under (\ref{c-t}). 

The situation is less clear when additional field and coupling constant redefinitions are invoked. In this latter case it has been stated in the literature \cite{cho-prl-92} that the action (\ref{bd-action}) is invariant under (\ref{c-t}), plus the following redefinitions (see also \cite{faraoni-rev-1999,faraoni-omega-infty,appendix}):

\be \bar\vphi=\vphi-2\ln\Omega,\;\;\bar\omega=\frac{\omega+6\der_\vphi\ln\Omega(1-\der_\vphi\ln\Omega)}{(1-2\der_\vphi\ln\Omega)^2}.\lab{f-red}\ee Contrary to existing claims it can be shown that, as a matter of fact, vacuum BD theory is not invariant under the transformations (\ref{c-t}), (\ref{f-red}), unless $\omega\neq -3/2$. Actually, even if it is true that the BD vacuum action (\ref{bd-action}) is invariant under (\ref{c-t}), (\ref{f-red}), this does not imply that the field equations of the theory are also invariant under these transformations. To demonstrate this it suffices to write the vacuum KG equation (\ref{v-bd-kg-eq}) in the equivalent form: $$\Box\vphi+\frac{1}{2}(\n\vphi)^2+\frac{R}{2\omega}=0.$$ This equation is clearly not conformally invariant, since the only conformal-invariant vacuum KG equation is necessarily of the form (note that this corresponds to the choice $\omega=-3/2$ in the last equation), $$\Box\vphi+\frac{1}{2}(\n\vphi)^2-\frac{R}{3}=0,$$ which can also be written in more standard notation \cite{faraoni-rev-1999,faraoni-prd-2007} if make the replacement $\vphi\rightarrow 2\ln\xi$ $\Rightarrow\;\Box\xi-R\,\xi/6=0$. The difficulty with conformal invariance is originated from the transformation properties of the D'lambertian under (\ref{c-t}), (\ref{f-red}); 

\bea &&\Omega^{-2}\Box\vphi=\frac{\bar\Box\bar\vphi}{1-2\der_\vphi\ln\Omega}+\nonumber\\
&&\;\;\;\;\;\;\;\;\;\;\;\;\;\;2\left[\frac{\der^2_\vphi\ln\Omega-\der_\vphi\ln\Omega(1-2\der_\vphi\ln\Omega)}{(1-2\der_\vphi\ln\Omega)^3}\right](\bar\n\bar\vphi)^2,\nonumber\eea so that the KG equation (\ref{v-bd-kg-eq}) transforms into: $$(2\bar\omega+3)\left\{\bar\Box\bar\vphi+\left[1+\frac{2\der^2_\vphi\ln\Omega}{(1-2\der_\vphi\ln\Omega)^2}\right](\bar\n\bar\vphi)^2\right\}=0.$$ This latter equation shares no resemblance with Eq. (\ref{v-bd-kg-eq}), unless, $\Omega^2=e^{2k\vphi}$, where the constant $k$ is a real number ($k\neq 1/2$). For the remaining gravitational field equations (Eq.(\ref{v-bd-feqs})), it is a matter of uncomplicated algebra to show that, unless $\omega=-3/2$, in which case (\ref{v-bd-feqs}) transform into, $$G_{\mu\nu}=-\frac{1}{2}[\der_\mu\vphi\der_\nu\vphi+\frac{1}{2}g_{\mu\nu}(\n\vphi)^2]+\n_\mu\n_\nu\vphi-g_{\mu\nu}\Box\vphi,$$ these are not invariant under (\ref{c-t}), (\ref{f-red}), neither. Since the vacuum BD action (\ref{bd-action}) is invariant under (\ref{c-t}), (\ref{f-red}), while, in general (arbitrary $\omega\neq -3/2$), the field equations derived from that action -- Eq.s (\ref{v-bd-feqs}), (\ref{v-bd-kg-eq}) -- are not invariant under these transformations, then the conformal invariance of vacuum BD theory is, at most, a mirage or spurious symmetry. As a result, the dynamics is different in the different conformal frames, i. e. the JF and EF of vacuum BD theory are not conformally equivalent representations just as it happens with Brans-Dicke theory with matter sources. 

A question then arises: what do authors who advocate physical (and/or mathematical) equivalence among JF and EF representations, actually understand by equivalence? Do they relate equivalence with an actual dynamical symmetry of the theory? We think this is a non enough explored aspect of the conformal transformation's issue that deserves being discussed. Most part of the misunderstanding arising within this context is due, precisely, to a lack of a clear definition of what to understand by ''equivalence''. To worsen things, we will show that there is an aspect of the issue that has not been discussed so far. It is connected with the possibility to assign a different geometrical meaning to conformal transformations of the metric. Aim of this paper is to provide new arguments that might help winning a dipper understanding of this thorny subject.

The novel aspect of the conformal transformation's issue we will uncover in the first part of this paper (mainly in sections \ref{map}, \ref{vp1-vs-vp2} and Sec.\ref{c-t-e-c}) is originated from an alternative interpretation of the conformal transformations of the metric, according to which, under (\ref{c-t}) not only the dynamical equations of BD theory are transformed but, simultaneously, the affine properties of space: the connection, the geodesics, etc, are also modified (a fact that is usually dismissed \cite{d-units,flanagan,faraoni-prd-2007,catena,capozziello-plb-2010,capozziello-mpla-2010,deruelle-sasaki,corda}). According to this approach, modification of the above mentioned affine properties of space is reflected in that, in terms of the original metric, for instance, the units of length may be point-independent -- as it is for Riemannian spaces -- while, in terms of the conformal metric, the length units may be point-dependent instead. Although the above fact has been partially considered in the seminal paper by Dicke \cite{d-units}, and more recently in \cite{faraoni-prd-2007} (see also \cite{catena}), it has passed unnoticed the fact that geometry with running (also changing or point-dependent) units of length can not be Riemannian but Weyl-integrable geometry\footnote{Here by ''geometry'' we understand the affine properties of space such as, the affine connection, the geodesics, etc.} instead \cite{quiros-2011}. In this understanding conformal transformations (\ref{c-t}) can be viewed as a mapping from Riemannian into Weyl-integrable spaces and vice versa, or, in other words, these can be properly understood as units transformations in the sense of Ref.\cite{d-units}. The resulting transformations of units relate complementary equally suited geometrical pictures of a same physical reality.

Another important aspect of the conformal transformation's issue that has to be carefully stated, no matter how obvious it seems, is to agree on what is to be meant by physical/mathematical equivalence among the different conformal frames in which a given gravity theory can be formulated. As discussed in the former page, unless a clear and mathematically meaningful statement of the latter notion is given, the conformal equivalence issue is no more than a semantic matter. In this paper, for definiteness, conformal equivalence is linked with invariance of the field equations under Weyl rescalings (see Eq.(\ref{scale-t}) below), which include conformal transformations (\ref{c-t}) plus a scalar field redefinition. Even if there can be other semantic uses of the word ''equivalence'', both from the mathematical and physical stand points it seems to us quite natural to associate conformal equivalence with conformal invariance. The latter property is comprised in what we call here as ''conformal equivalence principle'' (CEP), a clear statement of which is given in section \ref{cep}. Whether or not this is actually a fundamental principle of nature is not of importance to the results of this paper and will not be discussed here. In that section we shall explore a conformal-invariant theory as an example where the different conformal frames, in which the theory can be formulated, are actually (physically/mathematically) equivalent. It will be demonstrated that, in general, the mentioned equivalence can be achieved only within theories where the metric tensor is defined up to an (conformal) equivalence class of metrics. In the framework of such theories, unless a specific gauge is considered, the metric is not uniquely determined by the field equations. It is clear that BD theory does not belong in this group, since no specific gauge is necessary to get a closed system of equations determining the metric and the BD field: the Einstein-Brans-Dicke plus the Klein-Gordon equations form a closed system of equations in the $g_{\mu\nu},\vphi$ - variables. We will discuss this topic in detail in Sec.\ref{math-asp}.

In view of new arguments explored here: i) transformations (\ref{c-t}) can be understood as a mapping among Riemannian and Weyl-integrable spaces, and, ii) the notion of ''conformal equivalence'' is to be endowed with a concrete mathematical and physical meaning, a critical review of the different viewpoints existing in the bibliography on the issue is performed in Sec.\ref{debate}. Other relevant aspects of the conformal transformations conundrum, such as the positivity of the energy problem, as well as its implications for the singularity issue, will be discussed also (see sections \ref{c-t-e-c} and \ref{singularity-issue} respectively). Although in this paper, for simplicity, the discussion mostly relies on (vacuum) BD theory, the results of our study can be straightforwardly applied to scalar-tensor theories in general (see appendix B).\footnote{Here we deal with classical theories of the gravitational field. In consequence, arguments related with quantum properties and processes will not be considered.} In the next section, in order for the paper to be self-contained, the fundamentals of Weyl geometry (WG) are exposed. Weyl-integrable geometry (WIG) is a particular member in this latter class and is important for the novel understanding of the conformal transformations we shall reveal here.

\section{Fundamentals of Weyl geometry}\lab{weyl-geom}

Before we pursue the present discussion any further we do a step aside to expose the fundamentals of the simplest generalization of Riemannian geometry that is able to accommodate running units: Weyl geometry. For readable and pedagogical introduction to WG we recommend the classical books \cite{eddington}, \cite{pauli}, and \cite{adler}, however research papers can be found where the subject is exposed in a more or less pedagogical way \cite{bouvier,bib-weyl-geom,romero-ijmpa-2011,romero-2011}. It has to be said that, although there was a moderate revival of Weyl's ideas after Dirac's ''large numbers hypothesis'' \cite{dirac} (see also \cite{rosen}), in the last decades there has been a renewed interest in WG \cite{cheng,s-i-1,s-i-2,s-i-3,s-i-4,israelit,romero-ijmpa-2011,romero-2011,bd-area-metric}, in connection to the search for alternative explanations to outstanding questions of fundamental physics, such as the dark matter/dark energy issues. A Weyl space $({\cal M},g_{\mu\nu},w_\mu)$, is a manifold ${\cal M}$ endowed with a metric $g_{\mu\nu}$ and a (gauge) vector field $w_\mu$, so that the following ``metricity'' condition is satisfied:

\be \n^{(w)}_\mu g_{\alpha\beta}=-w_\mu g_{\alpha\beta},\lab{weyl-law}\ee where $\n^{(w)}_\mu$ is the Weyl covariant derivative operator, which is defined through the torsion-free affine connection of the Weyl space: 

\be \Gamma^{\;\alpha}_{\beta\gamma}=\{^{\;\alpha}_{\beta\gamma}\}+\frac{1}{2}\left(\delta^\alpha_\beta w_\gamma+\delta^\alpha_\gamma w_\beta-g_{\beta\gamma} w^\alpha\right),\lab{connection}\ee where 

\be \{^{\;\alpha}_{\beta\gamma}\}=\frac{1}{2}g^{\alpha\nu}\left(\der_\beta g_{\nu\gamma}+\der_\gamma g_{\nu\beta}-\der_\nu g_{\beta\gamma}\right),\lab{christoffel-s}\ee are the Christoffel symbols of the metric. For every (non-vanishing) differentiable function $\Omega$, the affine connection (\ref{connection}), and the metricity condition (\ref{weyl-law}), are invariant under the following transformations, also known as Weyl rescalings:

\be \bar g_{\mu\nu}=\Omega^2 g_{\mu\nu},\;\;\bar w_\mu=w_\mu-2\der_\mu\ln\Omega.\lab{transf}\ee Thus the metric $g_{\mu\nu}$, and the gauge vector $w_\mu$ are far from unique: rather $g_{\mu\nu}$ belongs in an equivalence class of metrics ${\bf g}$, so that, for each $g_{\mu\nu}\in{\bf g}$, there exists a unique gauge vector $w_\mu$, such that the law (\ref{weyl-law}) is satisfied. A given pair $(g_{\mu\nu},w_\mu)$ is called a gauge, and the transformations (\ref{transf}) are gauge transformations \cite{miritzis-cqg}. 

Due to (\ref{weyl-law}), under parallel transport, not only the orientation of a given vector changes, but, also its length $\ell=\sqrt{g_{\mu\nu}\ell^\mu\ell^\nu}$, varies from point to point in the Weyl manifold: $d\ell/\ell=dx^\mu w_\mu/2$. Hence, for instance, upon returning back to the starting point, after parallel transport in a closed path, the length of a vector will not be the same, $\ell=\ell_0\;\exp{\oint dx^\nu w_\nu/2}$. This feature of WG led Einstein to argue that electrons moving in a background of the $w_\mu$-field would produce unobserved broadening of the atomic spectral lines (see, however, arguments that overcome Einstein's objection \cite{cheng}). This broadening of the spectral lines is known as the ``second clock effect''.

\subsection{Weyl-integrable geometry}\lab{w-i-geom}

There is a particular subclass of WG which is free from the second clock effect; the so called Weyl-integrable (WI) geometry.\footnote{For applications of WIG in cosmology see, for instance, the review \cite{weyl-cosmo}, and also Ref.\cite{novello}.} The latter can be obtained from the general class of WG-s by replacing the gauge vector by a gradient of a scalar field: $w_\mu\rightarrow\der_\mu\vphi$. In general WI spaces can be represented by the triad $({\cal M},g_{\mu\nu},\vphi)$. In this case the length unit $\ell$ changes according to, $d\ell/\ell=dx^\mu\der_\mu\vphi/2=d\vphi/2$, so that, after parallel transport in a closed path, since $\oint d\vphi=0$, there is no neat change in the length unit $\oint d\ell/\ell=0$. Riemann spaces correspond to a particular gauge \cite{romero-ijmpa-2011} where $\vphi=\vphi_0=const\,\Rightarrow\,\nabla_\mu^{(w)}\rightarrow\nabla_\mu$, and the metric is convariantly constant: 

\be \nabla_\mu g_{\alpha\beta}=0,\lab{riemann-law}\ee where $\nabla_\mu$ refers to Riemannian covariant derivative operator defined through the Christoffel symbols. As a consequence, under parallel transport in a Riemannian space, vectors get rotated but their length is unchanged, i. e., length units are truly constant. 

An interesting feature of WI spaces is that the metricity condition (Eq.(\ref{weyl-law}) with the replacement $w_\mu\rightarrow\der_\mu\vphi$),

\be \n^{(w)}_\mu g_{\alpha\beta}=-\partial_\mu\vphi\;g_{\alpha\beta},\lab{wi-law}\ee the affine connection (Eq.(\ref{affine-c}) below), and several other geometric objects, are invariant under the following Weyl rescalings\footnote{In this paper sometimes we shall call as ''scale invariance'' invariance under the Weyl rescalings (\ref{scale-t}).}

\be \bar g_{\mu\nu}=\Omega^2 g_{\mu\nu},\;\;\bar\vphi=\vphi-2\ln\Omega.\lab{scale-t}\ee Hence, the metric $g_{\mu\nu}$ and the gauge scalar $\vphi$ are far from unique. Instead of a fixed pair $(g_{\mu\nu},\vphi)$ -- properly a gauge -- one has a whole (perhaps infinite) class of pairs 

\be {\cal C}=\left\{(g_{\mu\nu},\vphi)|\n^{(w)}_\mu g_{\alpha\beta}=-\der_\mu\vphi\,g_{\alpha\beta}\right\},\lab{c-class}\ee such that, any other pair $(\bar g_{\mu\nu},\bar\vphi)$ related with $(g_{\mu\nu},\vphi)$ by a Weyl rescaling (\ref{scale-t}), also belongs in ${\cal C}$. 

It is sometimes useful to write several geometric objects like, for instance, the WI curvature scalar $R^{(w)}$, Ricci tensor $R^{(w)}_{\mu\nu}$ and Einstein's tensor $G^{(w)}_{\mu\nu}$, respectively, in terms of their Riemannian counterparts:

\bea &&R^{(w)}=R-3\Box\vphi-\frac{3}{2}(\der\vphi)^2,\nonumber\\
&&R^{(w)}_{\mu\nu}=R_{\mu\nu}-\n_\mu\n_\nu\vphi-\frac{1}{2}g_{\mu\nu}\Box\vphi\nonumber\\
&&\;\;\;\;\;\;\;\;\;\;\;\;\;\;\;\;\;\;\;\;\;+\frac{1}{2}[\der_\mu\vphi\der_\nu\vphi-g_{\mu\nu}(\der\vphi)^2],\nonumber\\
&&G^{(w)}_{\mu\nu}=G_{\mu\nu}-\n_\mu\n_\nu\vphi+g_{\mu\nu}\Box\vphi\nonumber\\
&&\;\;\;\;\;\;\;\;\;\;\;\;\;\;\;\;\;\;\;\;\;+\frac{1}{2}[\der_\mu\vphi\der_\nu\vphi+\frac{1}{2}g_{\mu\nu}(\der\vphi)^2],\lab{weyl-riemann}\eea where, in the right-hand-side (RHS) of the above equations, stand usual Riemannian magnitudes, including the curvature scalar $R$, the Ricci tensor $R_{\mu\nu}$, the Einstein's tensor $G_{\mu\nu}=R_{\mu\nu}-g_{\mu\nu}R/2$, the covariant derivative operator $\n_\mu$, and the D'lambertian operator $\Box\equiv g^{\mu\nu}\n_\mu\n_\nu$, which are defined in terms of the Christoffel symbols (\ref{christoffel-s}). Weyl-integrable curvature quantities, instead, are defined in terms of the WI affine connection

\be \Gamma^{\;\alpha}_{\beta\gamma}=\{^{\;\alpha}_{\beta\gamma}\}+\frac{1}{2}\left(\delta^\alpha_\beta\der_\gamma\vphi+\delta^\alpha_\gamma\der_\beta\vphi
-g_{\beta\gamma}\der^\alpha\vphi\right).\lab{affine-c}\ee The WI Ricci and Einstein's tensors $R^{(w)}_{\mu\nu}$ and $G^{(w)}_{\mu\nu}$ are unchanged by the Weyl rescalings (\ref{scale-t}), while $\bar R^{(w)}=\Omega^{-2}R^{(w)}$, so that the scale-invariant measure of scalar curvature is the quantity, $e^{-\vphi}\;R^{(w)}$. Note, in between, that the quantity, $e^{\vphi/2}ds$, is a scale-invariant measure of spacetime separations. Other scale-invariant quantities of WIG are:

\be e^{-2\vphi}R^{(w)}_{\mu\nu}R_{(w)}^{\mu\nu},\;\text{and},\;e^{-4\vphi}R^{(w)}_{\alpha\beta\mu\nu}R_{(w)}^{\alpha\beta\mu\nu}.\label{sc-inv-q}\ee

Time-like geodesics in a WI space are described by the following scale-invariant equation \cite{bouvier}:

\be \frac{d}{ds}\left(\frac{dx^\alpha}{ds}\right)+\Gamma^\alpha_{\mu\nu}\frac{dx^\mu}{ds}\frac{dx^\nu}{ds}
-\frac{1}{2}\der_\mu\vphi\frac{dx^\mu}{ds}\frac{dx^\alpha}{ds}=0,\lab{wi-geod-eq}\ee where, as before, $\Gamma^\alpha_{\mu\nu}$ is the affine connection of the WI space (\ref{affine-c}), and the third term in the LHS of the equation is originated from variations of the units of length from point to point in the manifold. The latter term can be removed by an appropriate affine parametrization $\sigma=\sigma(s)$ $\Rightarrow d\sigma=e^{\vphi/2}ds$, so that the above geodesic equation can be rewritten in the standard way: 

\be\frac{d}{d\sigma}\left(\frac{dx^\alpha}{d\sigma}\right)+\Gamma^\alpha_{\mu\nu}\frac{dx^\mu}{d\sigma}\frac{dx^\nu}{d\sigma}=0.\lab{wi-geod-eq'}\ee The null geodesic equation is similar to Eq.(\ref{wi-geod-eq}) but with replacement of $ds\rightarrow d\lambda$, where $\lambda$ is an affine parameter along the null geodesic path. Unlike the standard case discussed in the bibliography (see, for instance \cite{faraoni-prd-2007,wald-appendix}), here the affine parameter $\lambda$ shares the same transformation properties with the interval $ds$ under (\ref{c-t}), namely $d\bar\lambda=\Omega\;d\lambda$.

\section{Two Faces of Conformal Transformations}\lab{map}

In this section we will explore an aspect of the conformal transformation's issue which has not been discussed so far. Transformations (\ref{c-t}) are usually understood as a mapping from Riemannian spaces into Riemannian spaces (first viewpoint below). Former studies of the issue have been performed under the implicit assumption that this interpretation of (\ref{c-t}) is the only possible. Here we will show that an alternative geometric interpretation of the conformal transformations is indeed possible (point of view exposed in subsection \ref{vp-2}). Let us assume we apply a transformation (\ref{c-t}) on the metric $g_{\mu\nu}$ of a Riemann's space. This means, in particular, that the connections of the starting manifold coincide with the Christoffel symbols of the metric (\ref{christoffel-s}), and, consequently, that the Riemann ``metricity'' condition (\ref{riemann-law}) is satisfied, resulting in that the length units in the starting space are point-independent. Under (\ref{c-t}) the Christoffel symbols transform as: 

\bea &&\left\{^\alpha_{\mu\nu}\right\}=\bar{\left\{^\alpha_{\mu\nu}\right\}}-\Omega^{-1}\left(\delta^\alpha_\mu\der_\nu\Omega\right.\nonumber\\
&&\left.\;\;\;\;\;\;\;\;\;\;\;\;\;\;\;\;\;\;\;\;\;\;\;\;\;\;\;\;\;\;+\delta^\alpha_\nu\der_\mu\Omega-\bar g_{\mu\nu}\bar g^{\alpha\sigma}\der_\sigma\Omega\right).\lab{christoffel-c-t}\eea If one compares this equation with Eq.(\ref{affine-c}), where the affine connection of a WI space is defined, one is left with two possibilities to build an affine structure into the conformal space.

\subsection{First point of view: Riemann$\mapsto$Riemann}\lab{vp-1} 

One possibility is just to regard the conformally related manifolds as endowed with different Riemannian structure of the same conformal class, so that Eq.(\ref{christoffel-c-t}) is just the transformation law relating $\left\{^\alpha_{\mu\nu}\right\}$ with $\bar{\left\{^\alpha_{\mu\nu}\right\}}$ under (\ref{c-t}). Assuming this interpretation -- the point of view adopted by most researchers in the field -- then the Riemannian metricity condition (\ref{riemann-law}) is unchanged, i. e. in the conformal space,

\be \bar\n_\mu \bar g_{\alpha\beta}=0,\lab{c-riemann-law}\ee so that the affine properties of space are not modified. In this case the transformation (\ref{c-t}) is just a mapping from Riemann's space into Riemann's space: 

\bea &&\gamma_{RR}:\text{Riemann}\mapsto\text{Riemann}\;\Leftrightarrow\nonumber\\
&&\;\;\;\;\;\;\;\;\;\,({\cal M},g_{\mu\nu})\mapsto({\cal M},\bar g_{\mu\nu}).\lab{rr}\eea 

Adopting this point of view amounts to consider that the units of length in the conformal space are point-independent as well. This is the most widespread viewpoint and it is not consistent with considering (\ref{c-t}) as a transformation of units in the sense of \cite{d-units}. Under this understanding, it happens, for instance, that time-like geodesics in the starting Riemannian space,

\be \frac{d}{ds}\left(\frac{dx^\alpha}{ds}\right)+\left\{^\alpha_{\mu\nu}\right\}\frac{dx^\mu}{ds}\frac{dx^\nu}{ds}=0,\lab{riemann-geod-eq}\ee are mapped into time-like curves which are not geodesics in the conformal (also Riemannian) space,\footnote{A different point of view on this property is exposed in \cite{faraoni-prd-2007} (see, however, the discussion in Sec.\ref{debate-eq}).}

\bea &&\frac{d}{d\bar s}\left(\frac{dx^\alpha}{d\bar s}\right)+\bar{\left\{^\alpha_{\mu\nu}\right\}}\frac{dx^\mu}{d\bar s}\frac{dx^\nu}{d\bar s}=\nonumber\\
&&\;\;\;\;\;\;\;\;\;\;\;\;\;\;\;\;\;\;\;\;\;\;\;\;\;\;\;\;\;\;\frac{\der_\mu\Omega}{\Omega}\left(\frac{dx^\mu}{d\bar s}\frac{dx^\alpha}{d\bar s}-\bar g^{\mu\alpha}\right).\lab{non-geod-eq}\eea Actually, it can be shown that, by means of an appropriate parametrization, one could remove the first term in the right-hand-side (RHS) of Eq. (\ref{non-geod-eq}) (see, for instance, Ref.\cite{wald}), however, the second term can not be eliminated \cite{faraoni-prd-2007}. In other words, the above equation does not admit an affine parametrization whatsoever, signaling a truly non-geodesic character of (\ref{non-geod-eq}). From the physical point of view, i. e., if identify the latter equation with the equation of motion of a test point-particle, the second term in the RHS can be identified with an additional force of non-gravitational origin acting on the test particle, commonly called ''five-force''. The latter happens to be an actual property of the laws of motion of a time-like particle in the conformal frame.

\subsection{Second point of view: Riemann$\mapsto$Weyl}\lab{vp-2}

The second possibility -- not explored so far in connection with the conformal transformation's issue -- can be consistently matched with the interpretation of (\ref{c-t}) as a transformation of units in the sense of Ref.\cite{d-units}. It is based on the following subtlety: take a second look at Eq.(\ref{christoffel-c-t}), and then, by comparing with (\ref{affine-c}), notice that one can safely identify the RHS of (\ref{christoffel-c-t}) with the definition of the affine connection, $$\bar\Gamma^{\;\alpha}_{\mu\nu}\equiv\bar{\{^{\;\alpha}_{\mu\nu}\}}-\Omega^{-1}\left(\delta^\alpha_\mu\der_\nu\Omega+\delta^\alpha_\nu\der_\mu\Omega-\bar g_{\mu\nu}\bar\der^\alpha\Omega\right),$$ of a conformal WI space $({\cal M},\bar g_{\mu\nu},\Omega)$. Then, under (\ref{c-t}), $\left\{^\alpha_{\mu\nu}\right\}\rightarrow\bar\Gamma^{\;\alpha}_{\mu\nu}$, $\n_\mu\rightarrow\bar\n^{(w)}_\mu$, so that the Riemannian metricity condition (\ref{riemann-law}) transforms into the WI metricity condition of the conformal space (compare with Eq.(\ref{c-riemann-law})):

\be \bar\n^{(w)}_\mu\bar g_{\alpha\beta}=2\Omega^{-1}\der_\mu\Omega\;\bar g_{\alpha\beta},\lab{c-w-i-law}\ee where the conformal factor $\Omega$ plays the role of the gauge scalar of the WI space.\footnote{Notice that if one suppresses the over bar and identifies $\vphi\equiv\ln\Omega^{-2}$, then, there is full resemblance with Eq.(\ref{wi-law}) of Sec.\ref{w-i-geom}.} According to this viewpoint, under the conformal transformation (\ref{c-t}), the original Riemannian space is mapped into a conformal WI space: 

\bea &&\gamma_{RW}:\text{Riemann}\mapsto\text{Weyl}\;\Leftrightarrow\nonumber\\
&&\;\;\;\;\;\;\;\;\;\;\,({\cal M},g_{\mu\nu})\mapsto({\cal M},\bar g_{\mu\nu},\Omega).\lab{rw}\eea That Eq.(\ref{c-w-i-law}) is not just a convenient rewriting of Eq.(\ref{c-riemann-law}) can be straightforwardly demonstrated. In the first place, notice that while the metricity condition (\ref{c-w-i-law}) is invariant under the following Weyl rescalings (compare with (\ref{scale-t})): $$\bar g_{\alpha\beta}\rightarrow\bar{\bar g}_{\alpha\beta}=\lambda^2\bar g_{\alpha\beta},\;\Omega\rightarrow\bar\Omega=\Omega-\lambda,$$ the Riemannian metricity condition (\ref{c-riemann-law}) does not obey this symmetry. The consequence is that, according to the first point of view displayed in Eq.(\ref{rr}), a unique metric tensor $\bar g_{\alpha\beta}$ is single out, meanwhile, according to Eq.(\ref{rw}) -- since, as mentioned, the WI metricity property (\ref{c-w-i-law}) is invariant under the above Weyl rescalings -- one is faced with a whole equivalence class of conformal metrics instead. In the second place, it is not difficult to prove that, according to (\ref{rw}), under (\ref{c-t}), Riemannian time-like geodesics (\ref{riemann-geod-eq}) are mapped into time-like geodesics of the conformal Weyl-integrable space:\footnote{Compare with WI geodesic equations given in Eq.(\ref{wi-geod-eq}). Complete resemblance with (\ref{wi-geod-eq}) is obtained if set, $\Omega^2=e^{-\vphi}$ $\Rightarrow\;\der_\mu\Omega/\Omega=-\der_\mu\vphi/2$.} 

\be \frac{d}{d\bar s}\left(\frac{dx^\alpha}{d\bar s}\right)+\bar\Gamma^\alpha_{\mu\nu}\frac{dx^\mu}{d\bar s}\frac{dx^\nu}{d\bar s}+\frac{\der_\mu\Omega}{\Omega}\frac{dx^\mu}{d\bar s}\frac{dx^\alpha}{d\bar s}=0,\lab{c-wi-geod-eq}\ee or, if choose an appropriate affine parametrization, $\bar\sigma=\bar\sigma(\bar s)$ $\Rightarrow d\bar\sigma=\Omega^{-1} d\bar s$, the above equation can be rewritten in the more standard way (see Sec.\ref{w-i-geom}): 

\be \frac{d}{d\bar\sigma}\left(\frac{dx^\alpha}{d\bar\sigma}\right)+\bar\Gamma^\alpha_{\mu\nu}\frac{dx^\mu}{d\bar\sigma}\frac{dx^\nu}{d\bar\sigma}=0.\lab{c-wi-geod-eq'}\ee This is to be contrasted with the usual understanding of the conformal transformations -- displayed by Eq.(\ref{rr}) -- according to which, Riemannian time-like geodesics (\ref{riemann-geod-eq}) are mapped into curves (\ref{non-geod-eq}), which do not admit an affine parametrization whatsoever and, hence, can not be geodesics (this is clearly demonstrated, for instance, in Ref.\cite{faraoni-prd-2007}). This subtlety and the resulting alternative interpretation of the conformal transformation (\ref{c-t}) displayed in Eq.(\ref{rw}), has not been explored before in connection with the conformal transformation's issue. The consequences of this novel aspect of the issue for gravity theories (BD theory in particular) is one of the subjects that will be investigated in the following sections.

\section{Conformal Transformations: Riemann$\mapsto$Riemann vs Riemann$\mapsto$Weyl}\lab{vp1-vs-vp2}

It is obvious that both interpretations of the conformal transformations (\ref{c-t}): $\gamma_{RR}$ (Eq.(\ref{rr})), and $\gamma_{RW}$ (Eq.(\ref{rw})), are mathematically correct, however, both have different geometrical (and physical) implications. In this section we will be assuming we deal with theories which are not conformally invariant, i. e. either the field equations are transformed by the conformal transformation (\ref{c-t}), which means that the laws of gravitation are different in the different conformal frames (first point of view), or, following the second viewpoint, the laws of gravity look the same but the geometrical interpretation is different in the different conformal spaces. Examples are BD theory and scalar-tensor theories of gravity (general relativity also belongs in this group).

\subsection{First viewpoint: physical implications}

According to the interpretation discussed in Sec.\ref{vp-1} (see Eq.(\ref{rr})), the transformations (\ref{c-t}) can be viewed as a mapping among Riemannian spaces, $\gamma_{RR}:({\cal M},g_{\mu\nu})\mapsto({\cal M},\bar g_{\mu\nu})$. Consequently, from the point of view of its geometrical implications, Eq.(\ref{c-t}) relates constant units of length in the starting space with constant units in the conformal space. The above interpretation of (\ref{c-t}) is not compatible with understanding conformal transformations of the metric as a transformation of units in the sense of \cite{d-units}. If we follow this viewpoint, since  geodesics of the starting Riemann's space are transformed into non-geodesic curves of the conformal, also Riemann's space, we have to regard (\ref{c-t}) as merely a mathematical transformation of the fields, relating spaces endowed with different Riemannian structure of the same conformal class, and with different dynamics on them. Physically this is illustrated by the fact that free falling test particles in the starting Riemannian space appear as particles which are not in free falling in the conformal (also Riemannian) space, a fact that can be explained as due to the presence of an additional non-gravitational interaction -- the so called ''five-force'' -- in the conformal variables. 

Under this reading of (\ref{c-t}), it is obvious that one is dealing with completely different theories with very dissimilar physical implications, which can be, in principle, tested by contrasting with the same observational data. Recall that the conformal transformations we are referring to in this paper are not coordinate transformations, i. e., the spacetime coincidences are not affected by the transformation (\ref{c-t}). Hence, if suitably refine the accurateness of the observational data-sets, in principle, a given conformal representation can be picked out. Take, as an example, vacuum Brans-Dicke theory of gravity (\ref{bd-action}), where the field equations which determine the dynamics are Eq.s (\ref{v-bd-feqs}), (\ref{v-bd-kg-eq}):

\bea &&G_{\mu\nu}=\omega\left[\der_\mu\vphi\der_\nu\vphi-\frac{1}{2}g_{\mu\nu}(\n\vphi)^2\right]+\der_\mu\vphi\der_\nu\vphi\nonumber\\
&&\;\;\;\;\;\;\;\;\;\;\;\;\;\;\;\;\;\;\;\;-g_{\mu\nu}(\n\vphi)^2+\n_\mu\n_\nu\vphi-g_{\mu\nu}\Box\vphi,\nonumber\\
&&\Box\vphi+(\n\vphi)^2=0.\lab{1}\eea Under (\ref{c-t}), according to $\gamma_{RR}$ in Eq.(\ref{rr}), the above equations transform into (\ref{efbd-feqs}), (\ref{efbd-kg-eq}):

\bea &&\bar G_{\mu\nu}=\left(\omega+\frac{3}{2}\right)\left[\bar\der_\mu\vphi\bar\der_\nu\vphi-\frac{1}{2}\bar g_{\mu\nu}(\bar\n\vphi)^2\right],\nonumber\\
&&\bar\Box\vphi=0.\lab{2}\eea Both sets of dynamical equations (\ref{1}) and (\ref{2}), are to be tied to Riemann's spaces $({\cal M},g_{\mu\nu})$, and $({\cal M},\bar g_{\mu\nu})$ respectively. Hence, JFBD and EFBD are just different theories related by a particular mathematical transformation Eq.(\ref{c-t}). In the first case -- JFBD theory given by equations (\ref{bd-action}), (\ref{1}) -- both fields, $g_{\mu\nu}$, and $\vphi$, determine the strength of gravity, while, in the second case -- EFBD theory depicted by Eqs.(\ref{ef-bd-action}), (\ref{2}) -- only the metric field propagates the gravitational interaction. The remaining scalar field $\vphi$, in this case, represents a matter source. Therefore, the (vacuum) EFBD theory is just general relativity with a (perhaps exotic) scalar matter source. If there were additional matter fields (other than the scalar), due to non-minimal coupling to the scalar field, these would interact non-gravitationally with $\vphi$. This additional interaction is what should be identified with the so called ''five-force''.\footnote{As we have shown, it is not necessary to have additional matter sources to reveal the existence of a five-force. Recall that, under the present viewpoint on (\ref{c-t}), the motion of test point particles in the conformal Riemannian space is non-geodesic.}

That the two conformally related theories are, in fact, different -- according to this point of view which is the most widespread -- is confirmed by the study of other issues such as positivity of energy which may be a principle of physics in one of the conformal representations while being violated in the other one (see the next section). Which one of the different conformally related theories is the ''physical'' one, is not a well-posed question in this understanding of the conformal transformations (\ref{c-t}), since each one of the conformal theories has its own set of measurable quantities -- mostly related to the invariants of the geometry -- so that these admit independent physical interpretation.\footnote{The corresponding measurable quantities can be contrasted with the experimental/observational evidence \cite{corda}. As a result, one can, at least in principle, experimentally/observationally differentiate among the different theories. While in the JF of BD theory one has to care about positivity of energy -- besides observational constraints from Solar system experiments which yield to unnaturally large values of the BD coupling $\omega>40000$ -- in the EFBD theory one has to meet the tight constraints coming from five-force experiments.} In this understanding the conformal transformations (\ref{c-t}) can serve, for instance, as a mathematical tool to deal with a set of differential equations, which are difficult to handle in the original variables, in terms of new field variables of the conformal space \cite{c-t-tool}. However, when extracting physical consequences one has to go back to the original field variables.

\subsection{Second viewpoint: physical implications}

The second viewpoint on the conformal transformations (see Eq.(\ref{rw})) is novel.\footnote{See, however, the reference \cite{romero-ijmpa-2011}, where a similar interpretation is managed through the concept of ''conformal Weyl frames''.} In this case the transformation (\ref{c-t}) relates Riemannian spaces with Weyl-integrable ones, $\gamma_{RW}:({\cal M},g_{\mu\nu})\mapsto({\cal M},\bar g_{\mu\nu},\Omega)$, i. e. constant units of length in the starting space are mapped into running units in the conformal space and vice versa. Under this alternative interpretation Eq.(\ref{c-t}) can be consistently understood as a units transformation in the sense of \cite{d-units}. In particular, geodesics of the starting Riemann's space are transformed into geodesics of the conformal WI space (see the discussion in Sec.\ref{vp-2}). It can be shown that, if one follows this point of view, then the gravitational laws -- expressed through the JFBD field equations (\ref{1}) -- will not be transformed. In this case what changes is the geometric interpretation of these laws. To show this, take as an example, vacuum BD theory in the Jordan's frame as the starting representation. Since, according to the viewpoint displayed by Eq.(\ref{rw}), under (\ref{c-t}), $$G_{\mu\nu}=\bar G^{(w)}_{\mu\nu},\;\n_\mu=\bar\n^{(w)}_\mu\;\Rightarrow\;\Box=\Omega^2\bar\Box_{(w)},$$ where $\bar\Box_{(w)}\equiv\bar g^{\mu\nu}\bar\n^{(w)}_\mu\bar\n^{(w)}_\nu$, then the vacuum JFBD field equations (\ref{1}) transform into,

\bea &&\bar G^{(w)}_{\mu\nu}=\omega\left[\bar\der_\mu\vphi\bar\der_\nu\vphi-\frac{1}{2}\bar g_{\mu\nu}(\bar\n\vphi)^2\right]+\bar\der_\mu\vphi\bar\der_\nu\vphi\nonumber\\
&&\;\;\;\;\;\;\;\;\;\;-\bar g_{\mu\nu}(\bar\n\vphi)^2+\bar\n^{(w)}_\mu\bar\n^{(w)}_\nu\vphi-\bar g_{\mu\nu}\bar\Box_{(w)}\vphi,\nonumber\\
&&\bar\Box_{(w)}\vphi+(\bar\n\vphi)^2=0,\lab{3}\eea i. e. the laws of gravity look the same in the conformal variables. However, since we start with a Riemannian space, while the transformations (\ref{c-t}) bring us into a conformal WI space, the geometrical interpretation of the measurements differs from one frame to the other one. In consequence, we are led with complementary geometrical representations of the same physics \cite{quiros}. These are equally suited and are not to be contrasted. 

It has to be pointed out that, in spite of the apparent invariance of the JFBD dynamics under units transformations (\ref{c-t}), one can not to talk about equivalence of the conformally related frames as associated with a dynamics preserving symmetry. Actually, if consider that, under (\ref{c-t}), $\gamma_{RW}:({\cal M},g_{\mu\nu})\mapsto({\cal M},\bar g_{\mu\nu},\Omega)$, then the JFBD action (\ref{bd-action}) is transformed into (recall that we are considering conformal transformations among the Einstein's and Jordan's frames so that $\Omega^2=e^\vphi$):

\be\bar S^{(w)}_{BD}=\frac{1}{16\pi}\int d^4x\sqrt{-\bar g}\left[\bar R^{(w)}-\omega (\bar\n\vphi)^2\right],\lab{wi-efbd-action}\ee where $\bar R^{(w)}$ is the WI curvature scalar given in terms of the conformal metric $\bar g_{\mu\nu}$, and so on. We can see that this action does not look the same as the original one in Eq.(\ref{bd-action}). In this case the question about which one of the conformal representations (frames) is the physical one does not arise. Both conformal frames are equally ''physical'' (or they are not). In particular, if positivity of energy holds in one representation, it will hold true also in the conformal frame (see the demonstration of this in the next section). The converse statement is also true. Additionally, since the field equations are unchanged under (\ref{c-t}), and, besides, these transformations leave unchanged the spacetime coincidences (properly the physical events), then observational testing can not differentiate between the conformally related representations of the theory.\footnote{Even if the JFBD and EFBD formulations are mathematically linked through a conformal transformation, $\bar g_{\mu\nu}=e^\vphi g_{\mu\nu}$, this does not mean that these representations are equivalent at all. Perhaps a closer notion could be ``duality'' or ``complementarity'' rather than ``equivalence''. In Ref.\cite{quiros}, for instance, the author relies on the notion of ``geometrical duality'' instead of ``conformal equivalence''. Duality of the conformal descriptions implies that these are different but mathematically related. Given that ''duality'' has been used in string theory in a quite different context \cite{wands}, complementarity can be a better suited synonym.}

\section{Transformations of units and positivity of energy}\lab{c-t-e-c}

Authors who refer to non-equivalence of the different conformal representations and favor, for instance, the Einstein's frame over the Jordan's one (see \cite{faraoni-rev-1999} for references), usually invoke arguments based on positivity of energy, and/or fulfillment of the energy conditions \cite{wald,hawking-ellis}. As long as one deals with the most widespread understanding of the conformal transformations (first point of view, subsection \ref{vp-1}), these arguments may be correct. However, if rely on the alternative viewpoint revealed in subsection \ref{vp-2} -- according to which under (\ref{c-t}) Riemann's spaces are mapped into WI ones and vice versa -- then these arguments may be wrong. 

According to the most widespread understanding of (\ref{c-t}), a first apparent argument against equivalence of the conformal frames comes from the straightforward comparison of JF and EF Brans-Dicke actions (\ref{bd-action}), and (\ref{ef-bd-action}), respectively, and is referred to positivity of the scalar field's kinetic energy. It is seen from EFBD action (\ref{ef-bd-action}) that, for $-3/2\leq\omega$, the kinetic energy of the scalar field has the correct sign while, for its conformal JFBD counterpart, provided that $\omega<0$, it shows the wrong sign. As an illustration, let us consider the graviton-dilaton sector of the string effective action in the so called string frame (SF). In this frame the graviton-dilaton action coincides with the JFBD action (\ref{bd-action}) with, $\omega=-1$, which implies that, $-3/2<\omega<0$. Hence, while in the SF graviton-dilaton effective string action the dilaton's kinetic energy has the wrong sign, in the conformal EF formulation the kinetic energy of the dilaton is positive definite instead. If, alternatively, invoke the second point of view displayed in Eq.(\ref{rw}) (subsection \ref{vp-2}), as it can be seen by comparing equations (\ref{bd-action}) and (\ref{wi-efbd-action}), the terms under squared brackets in the action are not transformed, so that the problem with non-positivity of the scalar field's kinetic energy in the original JFBD theory, is inherited by the alternative Weyl-integrable EF formulation. Needless to say that we are not considering here redefinition of the coupling constant $\omega$, which plays a similar role in both conformal formulations, providing the sign for the scalar field's kinetic energy. 

Similar arguments can be used to rule out statements found in the literature which point to apparent fulfillment of the energy conditions in one frame but not in the conformal one (see Ref.\cite{faraoni-rev-1999} and references therein). It is a well-known fact that the weak, strong and dominant energy conditions (WEC, SEC and DEC, respectively) \cite{wald,hawking-ellis} can all be violated by the scalar field $\vphi$ regarded as a form of matter in JF formulation of Brans-Dicke theory \cite{faraoni-prd-2007}. This is due to a term arising in the RHS of the JFBD field equations (\ref{v-bd-feqs}), which is linear in the second derivatives of $\vphi$, instead of being quadratic in the first derivatives. On the contrary, the Einstein's field equations (\ref{efbd-feqs}) -- which are derivable from the EFBD action (\ref{ef-bd-action}) -- are free of terms linear in the second derivatives of $\vphi$, so that there is no problem with fulfillment of the energy conditions in the EF formulation of BD gravity according to the first point of view in Sec.\ref{vp-1}. If consider, alternatively, the point of view displayed by Eq.(\ref{rw}), since the JFBD field equations are not transformed by (\ref{c-t}), then the terms linear in the second derivatives of the scalar field are preserved by the transformations of units. In this understanding of (\ref{c-t}), (non)fulfillment of the energy conditions in the original JF formulation of BD theory will entail (non)fulfillment of the energy conditions in the conformal EF representation of the theory.

\section{Conformal equivalence principle}\lab{cep}

What is called as ''conformal transformation's issue'' in the bibliography, is the apparent conundrum we face when trying to seek for an answer to the question: which one of the conformally related frames in which a given theory of gravity can be formulated is the physical one? This question makes sense only for those who understand that the different frames are not ''physically'' equivalent. Those who think the conformal frames are equivalent, obviously do not face this question. Due to the importance of a clear and meaningful statement of what is to be understood by ''equivalence'' in order to resolve the controversy, in this section we will endow the notion of ''conformal equivalence'' with a concrete physical/mathematical meaning. Recall that, otherwise, we will be facing a semantic debate, no more. We want to stress, however, that other precise statements of this notion can be possible (see, for instance, Ref.\cite{capozziello-mpla-2010}). Here we will follow the common sense and the notion of equivalence will be associated with a symmetry which preserves the dynamical content of the theory.

A concrete example where the meaning of the notion of ``equivalence'' is crystal clear is the famous Einstein's equivalence principle within special relativity (SR-EEP). The physical content of the SR-EEP can be stated in the following simple way: the laws of physics are the same no matter which one of the different inertial reference frames, in which these can be formulated, is chosen. Mathematically this means that there exists a set of linear (homogeneous) coordinate transformations -- Lorentz transformations -- which leave invariant, in particular, the differential equations that describe the given laws of physics. 

Following the above rule it is straightforward to formulate a principle of ``conformal equivalence'', or ``conformal equivalence principle'' (CEP for short), which might be a fundamental principle of nature whenever the laws of gravity are involved. From the point of view of its physical content, the CEP can be formulated in the following way: the laws of gravity look the same no matter which one of the different conformally related frames is chosen to describe them. From the mathematical point of view the CEP is to be associated with invariance of the field equations that describe the gravitational phenomena under the Weyl rescalings (\ref{scale-t}), which contain conformal transformations (\ref{c-t}). It is clear from the formulation of the CEP given above that physical conformal (non)equivalence implies mathematical conformal (non)equivalence and vice versa. In the framework of a conformal-invariant theory of gravity, for instance, all of the possible conformal frames in which the theory can be formulated are equally ``physical''. Even if the laws of gravity look simpler in one given conformal frame, none is preferred over the others. 

According to the above ''natural'' prescription, the statement about conformal equivalence of the different conformal frames will entail that the CEP is valid. The contrary statement is also true: if the CEP is not valid, then the different conformally related frames in which a given theory of gravity can be formulated are neither physically nor mathematically equivalent. As a particular example we may cite the Brans-Dicke theory of gravity. Since BD gravity theory is not conformally invariant -- the conformal transformation (\ref{c-t}) maps the JFBD field equations into the EFBD ones -- then the CEP is not valid. This means, in turn, that Einstein's and Jordan's conformal frames of Brans-Dicke theory are not equivalent. The argument can be safely applied to scalar-tensor theories in general. This example shows the importance of clearly prescribing what is meant by conformal equivalence for a meaningful discussion of the issue.

Although, as long as we know, it has not been as strictly formulated as we have done above, the principle of conformal equivalence has been assumed to play an important role in the understanding of the laws of physics in many influential papers before \cite{d-units,deser,waldron,waldron-tractors}. In this regard we want to point out that whether the CEP is a fundamental principle of nature is not a subject of interest in the present paper. In correspondence we will not make judgments about its validity here, so that the results of our discussion will not depend on the CEP being valid. Even in case it were a fundamental principle of physics, given the nature of the quantum measurement process, we do not expect the CEP to be valid at scales where quantum gravity effects become unavoidable. 

Nevertheless, it is of interest to investigate the physical and mathematical implications of a conformal-invariant theory of gravity. The latter represents a counterexample where, unlike BD theory, the CEP is satisfied. This is, precisely, the aim of the remaining part of this section. 

It is a matter of simple algebra to demonstrate that the particular value of the BD coupling, $\omega=-3/2$, is not transformed by (\ref{f-red}). Hence, the corresponding action,\footnote{That a scale-invariant scalar-tensor action corresponds to the (singular) choice $\omega=-3/2$ in BD theory has been shown, for instance, in the very well-known paper \cite{deser} (see also \cite{ruso1}).} $$S^{BD}_{3/2}=\frac{1}{16\pi}\int d^4x\sqrt{-g}\;e^\vphi\left[R+\frac{3}{2} (\n\vphi)^2\right],$$ together with the field equations derived from it, will be invariant under the Weyl rescalings (\ref{scale-t}). It makes sense, then, to rewrite the above action in terms of WI quantities by using the Riemannian decomposition of the WI curvature scalar $R^{(w)}$ in (\ref{weyl-riemann}). The result is:\footnote{In references \cite{romero-ijmpa-2011} and \cite{salim} a similar action was investigated in different contexts. In Ref.\cite{romero-ijmpa-2011}, besides the pure geometric part, also a ''cosmological constant'' and matter terms were considered.}

\be S^{(w)}=\frac{1}{16\pi}\int d^4x\sqrt{-g}\;e^\vphi\;R^{(w)}.\lab{bd-weyl}\ee In the present case not only the action (\ref{bd-weyl}) but also the field equations that can be derived from it,

\bea &&G^{(w)}_{\mu\nu}=0,\Rightarrow G_{\mu\nu}-\n_\mu\n_\nu\vphi+g_{\mu\nu}\Box\vphi\nonumber\\
&&\;\;\;\;\;\;\;\;\;\;\;\;\;\;\;\;\;\;\;\;\;+\frac{1}{2}[\der_\mu\vphi\der_\nu\vphi+\frac{1}{2}g_{\mu\nu}(\n\vphi)^2]=0,\nonumber\\
&&\;\;\;\;\;\;\;\;\;\;\;\;\;\;\;\;\;\;\;\;\;\Box\vphi+\frac{1}{2}(\n\vphi)^2-\frac{R}{3}=0,\lab{v-wi-bd-feq}\eea are invariant under the Weyl rescalings (\ref{scale-t}). It is evident that the CEP is satisfied in this formulation of the gravitational laws. The resulting conformal-invariant theory will be a fully geometrical description of the laws of gravity, not sharing any properties with the standard BD theory. In the particular gauge, $\vphi=\vphi_0$, when Riemannian geometry is recovered out of Weyl-integrable one, the action (\ref{bd-weyl}) is mapped into the standard Einstein-Hilbert action (see appendix A), $$S^{(w)}\rightarrow S_{EH}=\frac{1}{16\pi G_{eff}}\int d^4x\sqrt{-g}\;R,$$ where, as before, $R$ is the Riemannian curvature scalar, and, $G_{eff}=e^{-\vphi_0}$, is the effective gravitational coupling constant. I. e., in that gauge general relativity is recovered rather than Brans-Dicke theory. For that reason we may call the resulting scale-invariant theory of gravity as ''scale-invariant general relativity'' (see \cite{romero-ijmpa-2011}). 

In case the theory based on action (\ref{bd-weyl}) -- whose dynamics is dictated by the field equations (\ref{v-wi-bd-feq}) -- were a correct theory of gravity, the CEP were a fundamental principle of nature, so that, conformal symmetry were a true symmetry of the gravitational laws. This is to be contrasted with vacuum BD theory where this is a mirage or spurious symmetry instead. Put in different words: in the gravitational theory depicted by (\ref{bd-weyl}), (\ref{v-wi-bd-feq}), the laws of gravity look the same in the different conformal frames, while the spacetime coincidences -- properly the observations -- are unchanged. This is an outstanding example of a theory where, unlike BD theory, the different conformal descriptions of a given phenomenon are actually (physically and mathematically) equivalent.

\subsection{Positivity of energy}

After the analysis in section \ref{c-t-e-c}, it is almost trivial to show that the scale-invariant theory of gravity given by (\ref{bd-weyl}), (\ref{v-wi-bd-feq}), is free of the positivity of energy problem, including fulfillment of the energy conditions. In fact, the first thing to notice is that, in the action (\ref{bd-weyl}), the kinetic energy term of the gauge field $\vphi$ is absent. Hence there is no problem with positivity of the scalar field's kinetic energy. Besides, since the gravitational field equations are (see Eq.(\ref{v-wi-bd-feq})): $G^{(w)}_{\mu\nu}=0$, there are no terms linear in the second derivatives of the $\vphi$-field (in fact there are no terms containing derivatives of any order), so that there will be no problems with fulfillment of the energy conditions neither. This represents an additional difference between standard vacuum Brans-Dicke theory \cite{bd} and the conformal-invariant theory explored in this section.

\subsection{Other mathematical aspects}\lab{math-asp}

There are several mathematical consequences arising from assuming WI spaces as the natural geometrical basis for a scale-invariant theory of gravity (\ref{bd-weyl}), (\ref{v-wi-bd-feq}). In particular, due to invariance under (\ref{scale-t}), the metric $g_{\mu\nu}$ and the gauge scalar $\vphi$ are far from unique. Instead of a fixed pair $(g_{\mu\nu},\vphi)$, one has a whole (perhaps infinite) class of pairs ${\cal C}$ -- defined in Eq.(\ref{c-class}) -- such that, any other pair $(\bar g_{\mu\nu},\bar\vphi)$ related with $(g_{\mu\nu},\vphi)$ by a scale transformation (\ref{scale-t}), also belongs in ${\cal C}$. The situation is reminiscent of what happens when one invokes invariance under general coordinate transformations $\bar x^\alpha= f^\alpha(x^0,x^1,x^2,x^3)$: there is not a unique set of coordinates to describe a given physical situation, but a whole (in principle infinite) class of them.\footnote{We have to recall that invariance under diffeomorphisms and scale-invariance are independent symmetry requirements: conformal transformations of the kind considered here are not diffeomorphisms. It is evident that the scale-invariant theory given by the action (\ref{bd-weyl}) is also invariant under diffeomorphisms.} While, in the latter case, the spacetime coordinates are meaningless -- the physical meaning is transferred to the invariants of the geometry under spacetime diffeomorphisms -- in the case when conformal invariance is invoked, the fields themselves loss independent physical meaning. In this latter case the physically meaningful quantities are the conformal invariants of the WI space such as, for instance, the conformal-invariant measure of scalar curvature, $e^{-\vphi}\;R^{(w)}$, the conformal-invariant measure of spacetime separations, $e^{\vphi/2}ds$, as well as other WI conformal-invariant quantities given in Eq.(\ref{sc-inv-q}). 

The above discussed scale-invariance property is reflected in the mathematical structure of the field equations (\ref{v-wi-bd-feq}), which are derived from the action (\ref{bd-weyl}). In fact, when written in terms of Riemannian quantities, the first and second equations in the RHS of Eq.(\ref{v-wi-bd-feq}) are not independent from each other. The second equation is just the trace of the first one so that one equation is redundant. Hence, there will be 6 independent equations to determine 11 unknown degrees of freedom (10 components of the metric tensor plus the gauge field $\vphi$). Nonetheless, in addition to the 4 degrees of freedom to make diffeomorphisms -- four components of the metric can be transformed away -- one more component can be gauged away due to an additional degree of freedom to make scale transformations (\ref{scale-t}). In other words, up to general coordinate plus scale transformations, the field equations (\ref{v-wi-bd-feq}) ''uniquely'' determine the metric coefficients. Notice, however, that specifying a given $\vphi$, or, alternatively, one of the components of the metric tensor, amounts to going into a particular gauge. Hence, as a matter of fact, the field equations themselves are not enough to uniquely determine the field content of the theory. This ambiguity is a clear consequence of invariance under Weyl rescalings (\ref{scale-t}). 

To illustrate the point, let us, briefly consider vacuum cosmology within the context of the theory (\ref{bd-weyl}), (\ref{v-wi-bd-feq}). Assume a Friedmann-Robertson-Walker (FRW) spacetime with flat spatial sections, given by the line element $$ds^2=-dt^2+a^2(t)\delta_{ij}dx^idx^j,$$ where, as usual, $t$ is the cosmic time and $a(t)$ is the scale factor. The vacuum field equations (\ref{v-wi-bd-feq}) can then be written as follows ($H\equiv\dot a/a$, the dot accounts for $t$-derivative):

\bea &&3\left(H+\frac{1}{2}\dot\vphi\right)^2=0,\nonumber\\
&&\dot H+\frac{1}{2}\ddot\vphi-\frac{1}{2}\left(H+\frac{1}{2}\dot\vphi\right)\dot\vphi=0,\lab{v-cosmo-feq}\eea while the vacuum KG-equation (second equation in the RHS of Eq.(\ref{v-wi-bd-feq})): 

\be\dot H+2H^2+\frac{1}{2}\ddot\vphi+\frac{3}{2}\dot\vphi H+\frac{1}{4}\dot\vphi^2=0,\label{v-cosmo-kg-eq}\ee is not an independent equation. The Friedmann equation above can be integrated to obtain the following dependence of the scale factor upon the gauge field $\vphi$ ($a_0$ is an arbitrary integration constant):

\be a=a_0\;e^{-\vphi/2}.\lab{v-cosmo-sol}\ee If we substitute (\ref{v-cosmo-sol}) back into the remaining equations -- second equation in (\ref{v-cosmo-feq}) and equation (\ref{v-cosmo-kg-eq}) -- these become just identities ($0\equiv 0$), so that no new information can be extracted from them. This is a consequence of scale invariance since, due to this symmetry, we have the freedom to choose either any $\vphi(t)$, or any $a(t)$ we want. Recall that one of these degrees of freedom can be transformed in any desired way by an appropriate scale transformation of the kind (\ref{scale-t}). In particular, under the conformal transformation $\bar g_{\mu\nu}=e^\vphi g_{\mu\nu}$, the conformal scale factor, $\bar a=e^{\vphi/2} a(t)=a_0$, is a constant, as it should be in vacuum general relativity without cosmological constant. Consistently with this result, under the above transformation, the scale-invariant theory explored in this section transforms down into standard general relativity (see appendix A). 

This is, precisely, what one expects from a truly scale-invariant theory: either the gauge (scalar) field or one of the components of the metric tensor can be chosen at will, respecting, of course, mathematical consistency requirements such as the existence of the inverse conformal transformations, continuity, etc. (see \cite{deruelle-sasaki} for a similar discussion but in a non scale-invariant context). This freedom is what differentiates a truly conformally invariant theory from one which does not respect the CEP.

\section{Conformal equivalence: the debate}\lab{debate}

In this section we will examine the different approaches to the conformal transformation's issue encountered in the bibliography under the light of the new arguments given here: i) the conformal transformations (\ref{c-t}) admit, at least, two different geometrical interpretations (see sections \ref{map}, \ref{vp1-vs-vp2}), and, ii) in order to avoid the conformal transformations controversy being a semantic issue it is critical to agree on which concrete meaning is to be assigned under the notion of (physical/mathematical) ''equivalence''. In the present discussion, for definiteness, we will assume conformal equivalence to be associated with invariance under (\ref{scale-t}) as stated in section \ref{cep}. In consequence physical equivalence will entail mathematical equivalence and vice versa. 

Although the different approaches existing in the bibliography have been classified into several groups \cite{faraoni-rev-1999,magnano}, in the present discussion we will not follow the mentioned classification, and will consider only authors who adhere to one of the following two different types of approaches: i) those who consider that the JF and the EF are not equivalent \cite{faraoni-rev-1999,magnano,capozziello-plb-2010,corda,Nandi,Dick,Faraoni1,cauchy,Belluci,Jarv,Bhadar,Nozari,ruso2}, and ii) those who state that the JF and EF formulations of BD theory, and scalar-tensor theories in general, are equivalent \cite{d-units,faraoni-prd-2007,capozziello-mpla-2010,deruelle-sasaki,Alvarez,Fujii}.

\subsection{First Approach: JF and EF are not equivalent}\lab{debate-non-eq}

Authors in this group admit that the different conformally related frames, in particular Jordan's and Einstein's ones, are not physically equivalent (we refer the reader to the reviews \cite{magnano,faraoni-rev-1999} for a complete list of authors). In this sense, if follow the point of view exposed in subsection \ref{vp-1}, according to which the transformation (\ref{c-t}) is just a mapping $\gamma_{RR}:\text{Riemann}\mapsto\text{Riemann}$, we have to partially agree with these authors. In fact, if undertake this approach, since under (\ref{c-t}) the JFBD field equations (\ref{1}) are mapped into the EFBD ones (\ref{2}), both sets of equations represent different laws of gravity on Riemann space. Consequently, the different frames depict different theories of gravity with their own sets of measurable quantities (see the discussion in sections \ref{vp1-vs-vp2}, \ref{c-t-e-c}). 

Our disagreement arises when, according to these authors, it has to be cleared which one of the conformally related frames is the ''physical'' one, a problem which is properly known as the conformal transformation's issue. As it has been discussed in section \ref{vp1-vs-vp2}, the statement of this problem is not correct. In fact, since each frame represents a different theory with its proper dynamics, its own set of measurables, etc., the question about which one of the conformally related frames -- the JF and EF, in particular -- is the physical one, has to be replaced by a more pragmatic question: which one of the different conformal theories fits better the existing observational/experimental evidence? \cite{corda}. Besides, do not forget about first principles, such as positivity of energy, etc. which may also be checked.

\subsection{Second Approach: JF and EF are equivalent}\lab{debate-eq}

A second group of authors believe the different conformal frames are physically equivalent \cite{d-units,faraoni-prd-2007,capozziello-mpla-2010,deruelle-sasaki,Alvarez,Fujii} so that, in correspondence, the issue does not arise. In this case, the conformal transformation (\ref{c-t}) is understood as a units transformation. Here we will explore this approach in detail since, we think, there is a lot of confusion associated with the lack of a concise and mathematically definite statement of what the authors understand by ''equivalence''. In what follows, for definiteness, we will refer only to Jordan's and Einstein's conformal frames. 

In the famous paper \cite{d-units}, for instance, Dicke stated that ''...the laws of physics must be invariant under a transformation of units.'' No matter whether this statement is correct or not, the obvious fact is that Brans-Dicke theory itself does not belong in this class of theories. Hence, it is not understood which class of equivalence Dicke referred to in \cite{d-units}. In the well-known paper \cite{faraoni-prd-2007} -- which is fully consistent with Dicke's arguments -- the following statement is made: ''...the two frames are equivalent, provided that the units of mass, length, time, and quantities derived there from scale with appropriate powers of the conformal factor $\Omega$ in the Einstein frame.'' In the above quotation no clear statement is made neither of what to understand by ''equivalence''. Besides, in  concordance with Dicke's arguments, in Sec.III A of Ref.\cite{faraoni-prd-2007}, the authors say: ''Since physics is invariant under a change of units, it is invariant under a conformal transformation provided that the units of length, time, and mass...are scaled.'' What do the cited authors mean by ''invariance of physics'' under a conformal transformation?

Let us briefly revise the arguments given in Ref.\cite{faraoni-prd-2007}. In section III B, for instance, the authors study the motion of massive particles in the so called ''Einstein frame with running units''. They rely on the investigation of time-like geodesics. It is shown that the correction to the equation of motion -- see the RHS term of the EF non-geodesic equation (\ref{non-geod-eq}) -- is entirely due to variation of the particle's mass in the 3-space of an observer moving with the particle, so that, if consider that the mass of the particle in the EF varies as, $\bar m=\Omega^{-1} m$, due to running units, then there is not any effective modification of the geodesic motion. However, their argument is flawed since it is based on a wrong assumption, and, on a miss-interpretation of the role affine parametrization plays in geodesic motion. In the first place, the authors assume that, under (\ref{c-t}), the spacetime coordinates are also modified, $d\bar x^\mu=\Omega\;dx^\mu$.\footnote{See Eq.(3.1) of Ref.\cite{faraoni-prd-2007}, and compare with correct equations in Eq.(4) of the first entry in \cite{catena}, where $|ds|$ is involved rather than the $dx^\mu$.} Hence, according to their analysis, under (\ref{c-t}), the affine parameter along a time-like geodesic transforms like, $d\bar\lambda=\Omega^2 d\lambda$, which led them to come to the obviously wrong result that the line-element should transform like: $d\bar s^2=\Omega^4 ds^2$ (Eq.(3.16) of the cited paper). This later equation is clearly inconsistent with the original understanding that, under $\bar g_{\mu\nu}=\Omega^2 g_{\mu\nu}$, the spacetime separations $dx^\mu$ are unchanged \cite{d-units} ($d\bar x^\mu=dx^\mu$), which leads to: $d\bar s^2=\bar g_{\mu\nu}dx^\mu dx^\nu=\Omega^2 ds^2$, instead (see, for instance, Eq.(10) of reference \cite{d-units}). In the second place, the authors of \cite{faraoni-prd-2007} do not realize that the demonstration in Sec.III B of their paper, that the motion equations for EF worldlines do not admit an affine parametrization whatsoever, means, in fact, that, even assuming the EF mass changes as $\bar m=\Omega^{-1}m$, time-like motion in the Einstein's frame can not be geodesic at all (see our discussion of this matter in subsection \ref{vp-1}). Contrary to the conclusion extracted from this demonstration in \cite{faraoni-prd-2007}, as a matter of fact, this result would entail that the motion of time-like particles in the JF is not (dynamically) equivalent to the corresponding motion in the conformal ''EF with running units''. Actually, if understand the conformal transformations as they are usually considered (Sec.\ref{vp-1}), we see that, even allowing for mass units variation -- such as to make $\bar m=\Omega^{-1}m$ -- there is a neat modification of the time-like geodesics in the Einstein's frame. The demonstration is simple: just notice that the EF motion equation (\ref{non-geod-eq}) -- which is an alternative writing of equation (3.18) of Ref.\cite{faraoni-prd-2007} -- may be rewritten in the following form: $$\frac{d}{d\bar s}\left(\bar m\frac{dx^\alpha}{d\bar s}\right)+\bar m\bar{\left\{^\alpha_{\mu\nu}\right\}}\frac{dx^\mu}{d\bar s}\frac{dx^\nu}{d\bar s}=-\bar m\frac{\der_\mu\Omega}{\Omega}\bar g^{\mu\alpha},$$ where it has been considered that $\bar m=\Omega^{-1}m$ (the JF mass of the particle $m$ is a constant). If we compare the latter equation of motion with its JF counterpart: $$\frac{d}{ds}\left(m\frac{dx^\alpha}{ds}\right)+m{\left\{^\alpha_{\mu\nu}\right\}}\frac{dx^\mu}{ds}\frac{dx^\nu}{ds}=0,$$ it is seen, that even if consider that the EF masses scale as $\bar m=\Omega^{-1}m$, an additional term remains in the RHS of the EF motion equation, which can not be removed by an affine parametrization \cite{wald,carroll}, a fact that was demonstrated, precisely, in Sec.III B of \cite{faraoni-prd-2007}. Hence, the five-force effect can not be removed by allowing the units of length in the Einstein's frame to vary in the way it was considered in Ref.\cite{faraoni-prd-2007}. This shows that even if allow the units of length of the conformal frame to vary from point to point, the equations of motion are not the same in the conformally related frames, which means, in turn, that there is not any dynamical equivalence among the JF and the EF representations. 

In contrast to what is done in references \cite{d-units,faraoni-prd-2007}, a consistent consideration of the conformal transformations (\ref{c-t}) as transformations of units -- subsection \ref{vp-2} (see also Sec.\ref{vp1-vs-vp2} and \ref{c-t-e-c}) -- would yield to dynamical equivalence in the sense that, under (\ref{c-t}), JF time-like Riemannian geodesics are mapped into Weyl-integrable time-like geodesics (see the demonstration in subsection \ref{vp-2}). Hence, the missing argument in the analysis of references \cite{d-units}, \cite{faraoni-prd-2007} is the lack of consideration of the impact units transformations have on the modification of the affine properties of space (affine connection, geodesics, etc). Recall that in \cite{d-units,faraoni-prd-2007} it is implicitly assumed that both the starting and the conformal spaces are Riemannian in nature, so that, in particular, the time-like geodesics have to be those of a standard Riemannian metric. 

A very interesting approach that deserves independent comment is the one of Ref.\cite{catena}. In that reference the authors introduce frame-independent quantities and apply them to situations of cosmological interest. Besides, in the paper corresponding to the second entry in \cite{catena}, the authors study a frame invariant action (equation (7) of their paper). JFBD and EFBD theories correspond to particular gauges of their more general theory. While their analysis is correct, if regard the theory under consideration as a conformal-invariant theory, it is clear that these criteria can not be applied to Brans-Dicke theory, and scalar-tensor theories in general (see the discussion on this matter in the introduction). We think discussion of the affine properties of the underlying space (affine connection, geodesics, etc) is lacking in \cite{catena}. Notice that test particles in their theory do not follow geodesics of the metric $h_{\mu\nu}$ (here we use author's symbology). This is seen from the matter part of the action in Eq.(6) of \cite{catena}, where it is apparent that matter particles couple to the conformal geometry. This hints to possible five-force constraints on this theory.

\subsection{A third approach}

If adopt the viewpoint on the conformal transformations according to which, under (\ref{c-t}), $\gamma_{RW}:\text{Riemann}\mapsto\text{Weyl}$ $\Leftrightarrow({\cal M},g_{\mu\nu})\mapsto ({\cal M},\bar g_{\mu\nu},\Omega)$, i. e., Riemann's space (in the Jordan's frame variables) is mapped into a Weyl-integrable space (in Einstein's frame variables), then the point-dependent property of the units of length is already encoded in the affine structure of the conformal space. Besides, while the JF time-like geodesics are mapped into time-like geodesics of the conformal WI space (EF), the JFBD field equations (\ref{1}) are not transformed by (\ref{c-t}) ($\Omega^2=e^\vphi$). Hence, under this understanding of the conformal transformations (\ref{c-t}), the dynamics is unchanged, pointing to a kind of dynamical equivalence. Recall, however, that the geometric picture in the EF differs from the one in the JF -- in particular the action is modified by (\ref{c-t}) -- so that, in fact, what one actually has is two complementary geometrical descriptions of a same phenomenon. Additional arguments supporting this interpretation will be given in the next section, where it will be shown that, while in one frame one inevitably encounters spacetime singularities, in the conformal frame these singularities might be an infinite proper time into the future/past so that, in fact, these are removed from the alternative representation (see, also, a similar discussion in references \cite{quiros,qbc}).

\section{The singularity issue}\lab{singularity-issue}

In view of the possible impact of the developments presented in this paper, here we want to comment about a related very important subject: the singularity issue. Arguments in favor of the fulfillment of the energy conditions in one conformal frame but not in others, has led several authors to conclude that spacetime singularities in one frame might be avoided in a conformally related one \cite{quiros,kaloper,qbc}. These results have been criticized in Ref.\cite{faraoni-prd-2007} based on a re-analysis on the light of the so called ''Einstein's frame with running units'', which is in agreement with the spirit of Dicke's paper \cite{d-units}. According to \cite{faraoni-prd-2007}, since (following Dicke), the Jordan and Einstein frames are equivalent, singularities occur in the Einstein's frame if and only if they occur in the Jordan's frame. We have shown, however, that several arguments given in that reference are flawed, so that a new analysis of the subject is mandatory.

According to the alternative geometric interpretation of the conformal transformation (\ref{c-t}) given in Sec.\ref{vp-2} -- which is the one being geometrically consistent with Dicke's understanding of (\ref{c-t}) as a units transformation \cite{d-units} -- the study of the validity of the energy conditions, and, consequently, of the singularity issue, is not as trivial and straightforward as stated in \cite{quiros,kaloper,qbc,faraoni-prd-2007}. In fact, if we review the results of Sec.\ref{c-t-e-c} in accordance with the viewpoint displayed by Eq.(\ref{rw}), one immediately sees that, if the energy conditions \cite{wald,hawking-ellis} are not satisfied in one frame, these will not be met in its conformal frame either. The contrary statement is also true. This argument alone would support the results of the analysis in Ref.\cite{faraoni-prd-2007}, however, as already said, the situation is not so simple. Before reaching to any conclusive result one has to check, first, whether the singularity theorems in their standard formulation \cite{hawking-ellis} (which is given on the basis of Riemannian spacetimes), are valid when dealing with Weyl-integrable spaces. This issue deserves a separate investigation. The interesting thing is that, even without a strict formulation of the singularity theorems in WI spacetimes, one can reach to interesting qualitative results. 

To state the qualitative discussion on solid grounds, lets write the relationships between several curvature invariants in the JF and EF of BD gravity, under the assumption that the second point of view on (\ref{c-t}) -- Eq.(\ref{rw}) in subsection \ref{vp-2} -- is valid. Since, according to this alternative geometrical interpretation, $\gamma_{RW}:({\cal M},g_{\mu\nu})\mapsto({\cal M},\bar g_{\mu\nu},\Omega)$, then the following relationships are valid:

\be I_0=\Omega^2\bar I_0^{(w)},\;I_2=\Omega^4\bar I_2^{(w)},\;I_4=\Omega^8\bar I_4^{(w)},\lab{inv-rel}\ee where we have defined the following invariant quantities:

\be I_0\equiv R,\;I_2\equiv R^{\mu\nu}R_{\mu\nu},\;I_4\equiv R^{\mu\nu\sigma\rho}R_{\mu\nu\sigma\rho}.\lab{inv}\ee In Eq.(\ref{inv-rel}) the unbarred quantities refer to Riemannian invariants in JFBD variables, while the quantities with an over-bar are the ones given in the EF of BD theory coupled to WI spaces. Recall that the upper/lower index $(w)$ denotes the given quantity is given in terms of the affine connection of WIG, etc. 

Our analysis of the singularity issue will be based on the notion of geodesic (in)completeness, a concept which is independent of the affine properties of the spacetime manifold. In this regard we expect to show that in a given (conformal) formulation of the theory one can find ``sufficiently long'' time-like geodesics \cite{wald}, so that (time-like) geodesic incompleteness is not met and spacetime singularities that are present in one theory's formulation can be avoided in its conformal representation. 

For definiteness, let us suppose in EFBD theory there exists an isotropic spacetime singularity so that,\footnote{The study of the impact conformal transformations have on anisotropic singularities requires a more careful analysis.} following a time-like geodesic of the WI geometry, this singularity is necessarily met in a finite proper time into the future/past, $d\bar\tau\rightarrow\text{finite}$ ($\bar\tau$ is the proper time in EFBD variables). The above singularity is characterized by, say, $\bar I_4^{(w)}\rightarrow\infty$. According to (\ref{inv-rel}) -- assuming the singularity can be removed in the JFBD theory's representation -- one may find a function $\Omega$ such that, as the singularity is approached, $\Omega^8\rightarrow 0$, quick enough as to make the quantity, $I_4=\Omega^8\bar I_4^{(w)}\rightarrow\text{finite}$. This entails that, since under (\ref{c-t}), the elements of proper time in JF and EF of BD theory are related through, $d\tau^2=\Omega^{-2}d\bar\tau^2$, hence -- supposing the above assumptions are correct -- as the singularity is being approached, $d\tau\rightarrow\infty$. This means that in the Jordan's frame of the Brans-Dicke theory a singularity that is met in a finite proper time ($d\bar\tau\rightarrow\text{finite}$) in its conformally related EF, may be avoided in principle as long as $d\tau$ is large enough in the sense specified above \cite{wald}. Besides, divergent curvature invariants are mapped into finite ones and vice versa. 

We want to underline that, under the novel approach to (\ref{c-t}) explored in this paper, the point-dependent property of the units of length is already encoded in the affine structure of space (affine connection, geodesics, relevant curvature invariants, etc), so that, for instance, while in the EF, $\bar M=(d\bar\tau,\,\bar I_0^{(w)},\,\bar I_2^{(w)},\,\bar I_4^{(w)},...)$, is a set of measurable quantities in WI space, in the JF, $M=(d\tau,\,I_0,\,I_2,\,I_4,...)$, is the corresponding set of measurable quantities in Riemannian space. Therefore, as it is usually done, a straightforward analysis of the invariants will suffice to judge about the (non)occurrence of a curvature singularity in a given representation of BD theory. In other words, in contrast to the analysis of the singularity issue in Sec.IV of Ref.\cite{faraoni-prd-2007}, no additional ''machinery'' and/or technical assumptions have to be applied to the study of this subject.

Although there are subtle issues related with integrability (as a matter of fact we have to deal with integrals of the kind $\tau=\int d\bar\tau\,\Omega^{-1}$, rather than with infinitesimal quantities like $d\tau$) the above qualitative analysis is general enough. It shows that, for given functions $\Omega$, the singularities that are inherent in one frame of Brans-Dicke theory, may be avoided in a properly chosen conformal representation. For concrete examples where this happens we refer the reader to references \cite{quiros,kaloper,qbc,piao}. 

The lesson to be learned is that the singularity theorems \cite{wald,hawking-ellis}, which have been strictly formulated (and applied) in contexts that involve Riemannian spaces, are to be revised within the context of non-Riemannian spaces, WI ones being a particularly interesting case.

\section{Conclusion}\lab{conclusion}

In this paper we have re-examined the conformal transformation's issue on the light of new arguments: i) conformal transformations may be understood as a mapping among Riemannian and Weyl-integrable spaces (sections \ref{map}-\ref{c-t-e-c}), and, ii) the notion of ''conformal equivalence'' is to be endowed with a concrete mathematical and physical meaning (Sec.\ref{cep}). The main results of the present research can be summarized in the following way.

i) A novel aspect of the conformal transformation's issue has been revealed (subsection \ref{vp-2}). It is based on consideration of the effect transformations of units \cite{d-units} carry on the affine properties of space, such as the connection, the geodesics, etc.

ii) In the understanding that equivalence entails a dynamics preserving symmetry, we have shown that the JFBD and EFBD representations are not equivalent formulations of BD theory.

   \begin{itemize}

   \item If follow the most widespread understanding of the conformal transformations (\ref{c-t}) as a mapping, $\gamma_{RR}:\text{Riemann}\mapsto\text{Riemann}$, among Riemannian spaces, then, the JF and EF formulations of BD theory are actually different theories (with different dynamics) operating on Riemann spaces, each one with its own set of measurable quantities. Which theory is the most adequate one is a question that may be answered by the experiment. The question about which one of the conformal frames is the physical one is devoid of meaning.

   \item If follow the viewpoint explored in subsection \ref{vp-2}, according to which (\ref{c-t}) may be understood as a mapping, $\gamma_{RW}:\text{Riemann}\mapsto\text{Weyl}$, then the dynamics is the same in the JF and in the EF, however, the geometrical interpretation differs. Both conformal formulations of BD theory represent complementary geometrical descriptions of a same physics. The question about which one of the conformal frames is the physical one does not arise.
   
   \end{itemize}

iii) It has been shown that, contrary to claims existing in the bibliography \cite{cho-prl-92,faraoni-rev-1999,faraoni-omega-infty,appendix}, vacuum BD theory is not invariant under the conformal transformation (\ref{c-t}) plus the redefinitions (\ref{f-red}).

iv) We have explored an example of a theory whose conformal frames are actually equivalent representations of the theory. It has to be necessarily a scale-invariant theory \cite{romero-ijmpa-2011,salim} so that the metric is defined up to a class of conformal equivalence.

v) The main points of view existing in the bibliography on the conformal transformation's issue have been critically scrutinized. It has been shown, for instance, that several arguments in favor of equivalence \cite{faraoni-prd-2007} are flawed and fail to be meaningful due to lack of consideration of the effect units transformations carry on the affine properties of space. The singularity issue has been re-examined.

There remain several interesting questions to be investigated: Which is the actual geometrical/affine structure of spacetimes that serve as models of our universe? Are these Riemannian or non-Riemannian spaces? Is the CEP a fundamental principle of nature? If so were, is scale-invariant general relativity a better suited theoretical framework for the description of the classical laws of gravity than standard general relativity itself?

The authors thank Yun-Song Piao for pointing to us reference \cite{piao}, which might serve as an additional illustration of our discussion on the singularity issue. This work was partly supported by CONACyT M\'exico under grants 49865-F and I0101/131/07 C-234/07 of the Instituto Avanzado de Cosmologia (IAC) collaboration (http://www.iac.edu.mx/), by the Department of Physics, DCI, Guanajuato University, Campus Le\'on, and by the Department of Mathematics, CUCEI, Guadalajara University.

\section*{APPENDIX}

\subsection{General relativity gauge}\lab{gr-gauge}

General relativity can be obtained from (\ref{bd-weyl}) in a particular gauge when $\bar\vphi=\vphi_0$. In fact, after this choice, the solution of the equation in the RHS of Eq.(\ref{scale-t}) will be $$\vphi-\vphi_0=2\ln\Omega\;\Rightarrow\;\Omega^2=e^{\vphi-\vphi_0}.$$ It can be shown that, under (\ref{c-t}) with the latter choice of the conformal factor, the affine connection of a Weyl-integrable space maps to the Christoffel symbols of the conformal metric: $\Gamma^\alpha_{\beta\gamma}\rightarrow\bar{\{^{\;\alpha}_{\beta\gamma}\}}$, etc. In other words, WI spaces transform under (\ref{c-t}) -- with $\Omega^2=e^{\vphi-\vphi_0}$ -- into Riemannian spaces. Besides, the action (\ref{bd-weyl}) transforms into the Einstein-Hilbert action, $$S_{EH}=\frac{1}{16\pi G_{eff}}\int d^4x\sqrt{-\bar g}\;\bar R,$$ where $G_{eff}=e^{-\vphi_0}$ is the (rescaled) effective gravitational coupling. Hence GR is a particular gauge in the conformal equivalence class ${\cal C}$ defined in Eq.(\ref{c-class}). Going into this particular gauge means that conformal invariance becomes, automatically, into a broken symmetry.

\subsection{Brans-Dicke theory with matter}\lab{bd-matter}

Although for sake of simplicity we have considered in this paper only the vacuum sector of Brans-Dicke theory, several qualitative aspects of introducing matter into the theory can be discussed here. We will see that the results of the present work are not modified by the addition of ordinary matter into BD theory. We have seen, in particular, that considering the motion of test particles in vacuum BD theory already reveals several interesting features. For instance, assuming one adheres to the first viewpoint in subsection \ref{vp-1}, the test particles which follow geodesics of the JF metric do not fallow geodesics of the conformal EF metric. Hence, if one decides to interpret the results of a given analysis in the EF of BD theory, since deviations from geodesic motion are interpreted by the EF observers as due to the existence of additional interactions of non-gravitational origin between the test particle and the scalar field $\vphi$, one will be faced with confronting the predictions of the theory with ``five-force'' experiments. The above analysis may be corroborated by adding matter into Brans-Dicke theory. The matter part of the JFBD gravity is assumed to be given by \cite{bd}, $S_m=\int d^4x\sqrt{-g}{\cal L}_m(\psi_i,g_{\mu\nu})$, where ${\cal L}_m$ is the Lagrangian density of matter, and $\psi_i$ represent the matter fields. The following continuity equation is obeyed in the JFBD theory:

\be \n^\nu T^{(m)}_{\nu\mu}=0,\;T^{(m)}_{\mu\nu}=\frac{2}{\sqrt{-g}}\frac{\der}{\der g^{\mu\nu}}({\sqrt{-g}\;\cal L}_m),\lab{jf-cont}\ee where $T^{(m)}_{\mu\nu}$- the matter stress-energy tensor. Under (\ref{c-t}) with $\Omega^2=e^\vphi$ -- in the sense implied by the first viewpoint in Sec.\ref{vp-1} (Eq.(\ref{rr})) -- the JF matter action above is transformed into the EF one \cite{d-units,kaloper}, $\bar S_m=\int d^4x\sqrt{-\bar g}\;e^{-2\vphi}\bar{\cal L}_m(\bar\psi_i,e^{-\vphi}\bar g_{\mu\nu})$, while (\ref{jf-cont}) is mapped into 

\be \bar\n^\nu\bar T^{(m)}_{\nu\mu}=-\frac{1}{2}\bar\der_\mu\vphi\;\bar T_{(m)},\lab{ef-cont}\ee where we have considered that, under (\ref{c-t}), $\bar T^{(m)}_{\mu\nu}=\Omega^{-2}T^{(m)}_{\mu\nu}$, and $\bar T_{(m)}=\bar g^{\mu\nu}\bar T^{(m)}_{\mu\nu}$ is the trace of the EF stress-energy tensor. Only for traceless (massless) matter fields the continuity equation (\ref{jf-cont}) is not transformed by (\ref{c-t}). Physically Eq.(\ref{ef-cont}) means that in terms of the EFBD metric there is an additional (non-gravitational) interaction between matter and the scalar field (the five-force) so that these fields exchange energy-momentum. If one invokes instead the point of view of section \ref{vp-2}, $\gamma_{RW}:({\cal M},g_{\mu\nu})\mapsto({\cal M},\bar g_{\mu\nu},\vphi)$, then the JF continuity equation (\ref{jf-cont}) is mapped into 

\be \bar\n^\nu_{(w)}\bar T^{(m)}_{\nu\mu}=-\bar\der^\nu\vphi\;\bar T^{(m)}_{\nu\mu},\lab{wi-cont}\ee which is just the continuity equation in WI spaces. Although the term in the RHS of (\ref{wi-cont}) might seem alien, it expresses the fact that the units of measure of the stresses and energy are running units. In no case the RHS of (\ref{wi-cont}) can be interpreted as a source term. To show this notice that Eq.(\ref{wi-cont}) is the trace of the most general equation, $\bar\n^{(w)}_\alpha\bar T^{(m)}_{\nu\mu}=-\bar\der_\alpha\vphi\;\bar T^{(m)}_{\nu\mu}$, which is, in turn, the equivalent of the metricity condition (\ref{c-w-i-law}) -- with $\Omega^2=e^\vphi$ -- in the matter sector of the EFBD theory. The opposite sign in the RHS of the above equation in respect to the sign of the RHS of Eq.(\ref{c-w-i-law}) is a consequence of the fact that, increasing extent of the units of length and time is correlated with the contrary effect on the units of stresses and energy.

\end{document}